\documentstyle[prb,12pt,aps,preprint]{revtex}
\draft
\tighten
\def\varlatex{{\rm L\kern-.36em\raise.3ex\hbox{A}\kern-.15em
    T\kern-.1667em\lower.7ex\hbox{E}\kern-.125emX}}
\global\firstfigfalse
\global\firsttabfalse
\begin{document}
\language=0
\input{psfig}
\def\bild#1#2#3#4#5#6{\centerline{\hbox{\psfig{figure=#1,height=#2
,bbllx=#3bp,bblly=#4bp,bburx=#5bp,bbury=#6bp,clip=}} }}
\def\ket#1{\vert\, #1 \, >}
\def\bra#1{< \, #1 \, }
\def\acct#1{#1^\prime }
\def\facu#1#2{\left( \begin{array}{c} #1 \\ #2 \end{array} \right)}
%
%
\def\slash#1{\setbox0=\hbox{$#1$}\setbox1=\hbox{$/$}\if\wd0>\wd1
\copy0\kern-\wd0\hbox to
\wd0{\hss\kern.1em\box1\hss}\else\copy1\kern-\wd1\hbox to
\wd1{\hss\box0\kern.1em\hss}\fi}
%
%
\font\smallcaps=cmcsc10 scaled\magstep1
\font\bfbig=cmbx10 scaled\magstep1
%
%
\title{EVOLUTION EFFECTS ON THE NUCLEON DISTRIBUTION AMPLITUDE}
\vspace{-0.6 true cm}
\author{
M. Bergmann\thanks{Invited talk presented at the Workshop on Exclusive
Reactions at high Momentum Transfer, Elba, Italy, 24-26 June, 1993, to
be published in the Proceedings.}
and N. G. Stefanis}
\address{Institut f\"ur Theoretische Physik II \\
Ruhr-Universit\"at Bochum, D-44780 Bochum, Germany\\
E-mail: michaelb@hadron.tp2.ruhr-uni-bochum.de}
\maketitle
\vspace{-0.5 true cm}
\begin{abstract}
We study the Brodsky-Lepage evolution equation for the nucleon and
construct an eigenfunction basis by including contributions of up to
polynomial order $9$. By exployting the permutation symmetry
$
P_{13}
$
of these eigenfunctions, a basis of symmetrized Appell polynomials
can be constructed in which the diagonalization of the evolution kernel
is considerably simplified. The anomalous dimensions are calculated
and found to follow a power-law behavior. As an application, we consider
the Brodsky-Huang-Lepage ansatz.
An algorithm is developed to properly incorporate such higher order
contributions in a systematic way.
\end{abstract}
\vspace{-0.2 true cm}
\section{General framework}
\vspace{-0.4 true cm}
The momentum scale dependence of the nucleon distribution
amplitude is given by the
Brodsky-Lepage evolution equation~\cite{LB80}
\begin{equation}
x_1 x_2 x_3 \left[ \frac{\partial}{\partial \xi}
\tilde{\Phi}_i(x_i,\xi)\, +\,
\frac{3}{2}\frac{C_F}{\beta} \tilde{\Phi}(x_i,\xi) \right] =
\frac{C_B}{\beta}
\int_{0}^{1} [dy] V[x_i,y_i] \tilde{\Phi}(y_i,\xi),
\label{evgl}
\end{equation}
with $\xi = \ln{\ln{\frac{Q^2}{\Lambda_{QCD}^2}}}$,
$C_B = \frac{N_c+1}{2N_c} = \frac{2}{3}$ and
$C_F = \frac{N_c^2-1}{2N_c} = \frac{4}{3}$ the bosonic and fermionic
Casimir-operators of $SU(N)_{color}$, respectively;
$\beta = 11-\frac{2}{3}\, N_F=9$ being the Gell-Mann and Low function.
The asymptotic solution of this equation is
$\Phi_{AS}\, = \, 120\,x_1x_2x_3$.
To leading order in $\alpha_{s}$, the interaction kernel between quark
pairs $\{i,j\}$ is
given in Ref.~\onlinecite{LB80}.
Factorization of Eq.~(\ref{evgl}) leads to
{\footnotesize
\begin{equation}
-\eta\,\, x_1 x_2 x_3 \,\tilde{\Phi}(x_i,\xi)  =
\int_{0}^{1} [dy] V[x_i,y_i] \tilde{\Phi}(y_i,\xi)\ \ \text{and}\ \
x_1 x_2 x_3 \left[ \frac{\partial}{\partial \xi}
\tilde{\Phi}(x_i,\xi)\, +\, \gamma\tilde{\Phi}(x_i,\xi) \right] =  0
\end{equation}
}
with  $\gamma=\frac{3}{2}\frac{C_F}{\beta} + \eta \frac{C_B}{\beta}$
and the integration measure is defined by
$\int_0^1 [dx] \equiv \int_0^1 dx_1 \int_0^{1-x_1} dx_2$\ $\int_0^1 dx_3
\ \delta (1-x_1-x_2-x_3)$.
[In the following we use the operator
$\int_{0}^{1} [dy]\, V[x_i,y_i]  \mapsto
\hat{V}$, which commutes with the permutation operator
$\hat{P}_{13}\ \ (x_{1}\Leftrightarrow x_{3})$.]
The evolution behavior of the eigenfunctions comes from the sensitivity
on the transverse momentum integration~\cite{LB80} arising from the
gluon exchange kernels. It can be expressed in the form
$\tilde{\Phi}(x_i,Q^2) = \ln^{-\gamma_{F}}{\left(
\frac{Q^2}{\Lambda_{QCD}^2}
\right) } \, \,
\tilde{\Phi}(x_i)$.
The representation of the evolution kernel is conveniently described in
terms of Appell polynomials,
$\tilde{{\cal F}}_{mn}(x_1,x_3)$,
which constitute an orthogonal basis with
weight~\cite{LB80} $\Phi_{AS}(x_i)/120$.
It was shown in Ref.~\onlinecite{Tes82} that in this basis, $\hat{V}$
becomes block-diagonal with respect to different polynomial orders
$m+n$.
Because of
$[\hat{P}_{13},\hat{V}]=0 $,
it is useful~\cite{Ber93,BS93} to define a symmetrized basis
of Appell polynomials
\begin{equation}
 \tilde{{\cal F}}_{mn}(x_1,x_3) = (1/2)\,\left(
 {\cal F}_{mn}(x_1,x_3) \pm {\cal F}_{nm}(x_1,x_3) \right)
\quad \text{for} \ (m \ge n\ / \ m<n).
\end{equation}
The particular importance of this basis lies in the fact that
$\hat{V}$ is block-diagonal within a particular order for different
symmetry classes ($S_n=\pm 1$) with respect to $\hat{P}_{13}$.
As a result, it is possible to analytically diagonalize $\hat{V}$
up to order $7$ (cf. Tab.~\ref{Tpoly}) [up to order
$({\cal O}(n)=)\, M=9$ this is done in Refs. 3,4].
The eigenvalues of any order ${\cal O}(n)$ and for a specific symmetry
class $S_n$ form a ``multiplet''-like set (see Fig.~\ref{fig1}),
which follows an approximate power-law:
$\gamma_n = 0.37 \, {\cal O}(n)^{0.565}$ that
differs from that of scalar (S=0) and vector (S=1)
mesons~\cite{Ber93,BS93}. This observation is in contrast
to the claims of Ref.~\onlinecite{EHG92}. The eigenfunctions of
the evolution equation have properties of a commutative group
\vspace{-0.2 true cm}
\begin{equation} \label{strukprod}
\tilde{\Phi}_k(x_i) \, \tilde{\Phi}_n(x_i)
 = \sum_{l=0}^\infty F_{kn}^l \tilde{\Phi}_l(x_i) \ \
\text{with} \ \ \vert {\cal O}(k)-{\cal O}(n) \vert \, \leq \,
{\cal O}(l) \, \leq \, {\cal O}(k)+{\cal O}(n)
\end{equation}
and obey the above triangle relation.
The structure coefficients~\cite{Ber93,BS93} $F_{kn}^l$ of the group are
calculated by $
 F_{kn}^l = N_l\,\int_0^1 [dx]\, x_1x_3(1-x_1-x_3)\,\tilde{\Phi}_k(x_i)
                 \tilde{\Phi}_n(x_i) \tilde{\Phi}_l(x_i)$,
with the particularly important case $ F_{kk}^0 = \frac{N_0}{N_k}$.
\vspace{-0.6 true cm}
\section{APPLICATIONS}
\vspace{-0.4 true cm}
The evolution effect of the nucleon distribution amplitude is important
for the calculation of various form factors at intermediate values
of the momentum transfer $Q^2\approx 10\div 30\, GeV^2/c^2$ (see
Ref.~\onlinecite{LB80,Ber93,BS93,EHG92}). In this
range, $\alpha_S(Q^2)$ tends to diverge whereas the $Q^2$ evolution of
the distribution amplitude, given by
\begin{equation}
\Phi_N(x_i,Q^2) = \Phi_{AS}(x_i) \left( \sum_{n=0}^{n_{max}} B_n(Q^2)
\tilde{\Phi}_n(x_i) \right), \label{ansz}
\end{equation}
significantly reduces the value of the form factors by more than
$\approx 30\%$. In the above representation of $\Phi_N$, contributions
of eigenfunctions up to polynomial order 3 have been
studied~\cite{EHG92,Sch89}. By incorporating on the rhs of eq.
(\ref{ansz}) (c.f. Ref.~\onlinecite{HS91})
the factor
$ f(x_i,\lambda_j) = e^{{-\lambda_1^2\left( \sum_{i=1}^3
\frac{1}{x_i} -\lambda_2^2 \right)}}$,
eigenfunctions of higher orders can be taken into account
having recourse to a Brodsky-Huang-Lepage type of ansatz.
This extended ansatz for $\lambda_1 = 0.03$ and $\lambda_2 = 3$ has
been studied in Ref.~\onlinecite{EHG92}. In order to determine its evolution
behavior, one has to project~\cite{HS91} $\Phi_N$ on the eigenfunctions
$\tilde{\Phi}_n$. For this purpose, $f$ can be expanded in the basis of
eigenfunctions of the nucleon evolution equation
\begin{equation}
f(x_i,\lambda_j) = \sum_k c_k(\lambda_j)\,\tilde{\Phi}_k(x_i), \ \
\text{with}\ \ c_k(\lambda_j) = N_k\,\int_0^1 [dx]\,x_1x_2x_3\,
f(x_i,\lambda_j) \tilde{\Phi}_k(x_i).
\end{equation}
This expansion and the properties of products of eigenfunctions given
by the structure coefficients $F_{nk}^l$ lead to
\vspace{-0.4 cm}
\begin{equation}
\Phi_N(x_i) = \Phi_{AS}(x_i)  \left( \sum_l \hat{B}_l\,
                \tilde{\Phi}_l(x_i) \right) \ \
\text{with}\ \  \hat{B}_l\, = \left( \sum_{n,k} B_n c_k F_{nk}^l \right).
\end{equation}
In the last equation the evolution of the distribution amplitude
is fully determined by the scale-dependence of the expansion
coefficients
$\hat{B}_l(Q^2) = \hat{B}_l(\mu^2)
\left( \frac{\alpha_S(Q^2)} {\alpha_S(\mu^2)} \right)^{\gamma_l}$,
which are a complicated mixture of the original
expansion coefficients $B_n$ and the projection coefficients $c_k$
of the extended ansatz. In contrast to the assumption\cite{EHG92}
\vspace{-0.2 cm}
\begin{equation}
\Phi_N(x_i,Q^2) = \Phi_{AS}(x_i) \left( \sum_{n=0}^{n_{max}}
B_n(\mu^2)\,
\left(\frac{\alpha_S(Q^2)}{\alpha_S(\mu^2)}\right)^{\gamma_n}\,
\tilde{\Phi}_n(x_i) \right) f(x_i,\lambda_j),
\end{equation}
the evolution of the distribution amplitude is {\it
not} found to be determined by the scale-dependence
$B_n(Q^2) = B_n(\mu^2) \left( \frac{\alpha_S(Q^2)}
{\alpha_S(\mu^2)} \right)^{\gamma_n}$. The deviation of the
approximation~\cite{EHG92} from the correct evolution behavior
can be expressed by the Q-dependent ratio
\vspace{-0.2 cm}
\begin{equation}
{\cal M}_l(Q^2) =
{\left(
{\hat{B}_l(\mu^2)\,\left(\frac{\alpha_S(Q^2)}{\alpha_S(\mu^2)}
\right)^{\gamma_l}}\right) }{  / } {\left(
{\sum_{n,k} B_n(\mu^2) \left(\frac{\alpha_S(Q^2)}{\alpha_S(\mu^2)}
\right)^{\gamma_n}
 c_k F_{nk}^l}\right)}.
\end{equation}
\vspace{-0.8 true cm}
\begin{table}
\squeezetable
\caption{ Orthogonal eigenfunctions
$\tilde{\Phi}_n(x_1,x_2,x_3)=\sum_{lk}\, a_{kl}^{n}\, x_1^kx_3^l$
of the nucleon evolution equation (represented by the
coefficient matrix $a_{kl}^{n}$ with $a_{kl}^{n} = S_n \, a_{lk}^{n}$;
$a_{22}^{n}=0$ for all n). The normalization is given by
$\int_0^1\, [dx]\, x_1x_2x_3\, \tilde{\Phi}_k(x_i) \tilde{\Phi}_n(x_i) =
(N_n)^{-1}\,  \delta_{kn}$. }
\label{Tpoly}
\begin{tabular}{rc|rccc}
$ n  $&$ M $&$ S_n $&$\gamma_n$&$\eta_n$&$ N_n$ \\
\hline
 $ 0$&$ 0$&$1$&${2\over {27}}$&$-1$&$120$\\
 $ 1$&$ 1$&$-1$&${{26}\over {81}}$&${2\over 3}$&$1260$\\
 $ 2$&$ 1$&$1$&${{10}\over {27}}$&$1$&$420$\\
 $ 3$&$ 2$&$1$&${{38}\over {81}}$&${5\over 3}$&$756$\\
 $ 4$&$ 2$&$-1$&${{46}\over {81}}$&${7\over 3}$&$34020$\\
 $ 5$&$ 2$&$1$&${{16}\over {27}}$&${5\over 2}$&$1944$\\
 $ 6$&$ 3$&$1$&${{115 - {\sqrt{97}}}\over {162}}$&${{-\left( -79 + {\sqrt{97}}
\right) }\over {24}}$&${{4620\,\left( 485 + 11\,{\sqrt{97}} \right) }\over
{97}}$\\
 $ 7$&$ 3$&$1$&${{115 + {\sqrt{97}}}\over {162}}$&${{79 + {\sqrt{97}}}\over
{24}}$&${{4620\,\left( 485 - 11\,{\sqrt{97}} \right) }\over {97}}$\\
 $ 8$&$ 3$&$-1$&${{559 - {\sqrt{4801}}}\over {810}}$&${{-\left( -379 +
{\sqrt{4801}} \right) }\over {120}}$&${{27720\,\left( 33607 -
247\,{\sqrt{4801}} \right) }\over {4801}}$\\
 $ 9$&$ 3$&$-1$&${{559 + {\sqrt{4801}}}\over {810}}$&${{379 +
{\sqrt{4801}}}\over {120}}$&${{27720\,\left( 33607 + 247\,{\sqrt{4801}} \right)
}\over {4801}}$\\
 $ 10$&$ 4$&$-1$&${{346 - {\sqrt{1081}}}\over {405}}$&${{-\left( -256 +
{\sqrt{1081}} \right) }\over {60}}$&${{196560\,\left( 7567 - 13\,{\sqrt{1081}}
\right) }\over {1081}}$\\
 $ 11$&$ 4$&$-1$&${{346 + {\sqrt{1081}}}\over {405}}$&${{256 +
{\sqrt{1081}}}\over {60}}$&${{196560\,\left( 7567 + 13\,{\sqrt{1081}} \right)
}\over {1081}}$
\end{tabular}
\vspace{-7 pt}
\begin{tabular}{r|cccccccc}
$ n  $&$a_{00}^n$&$a_{10}^n$&$a_{20}^n$&$a_{11}^n$&$a_{30}^n$&
       $a_{21}^n$&$a_{40}^n$&$a_{31}^n$ \\
\hline
 $  0$&$1$&$0$&$0$&$0$&$0$&$0$&$0$&$0$\\
 $  1$&$0$&$1$&$0$&$0$&$0$&$0$&$0$&$0$\\
 $  2$&$-2$&$3$&$0$&$0$&$0$&$0$&$0$&$0$\\
 $  3$&$2$&$-7$&$8$&$4$&$0$&$0$&$0$&$0$\\
 $  4$&$0$&$1$&$-{4\over 3}$&$0$&$0$&$0$&$0$&$0$\\
 $  5$&$2$&$-7$&${{14}\over 3}$&$14$&$0$&$0$&$0$&$0$\\
 $  6$&$1$&$-6$&${{41 + {\sqrt{97}}}\over 4}$&${{3\,\left( 31 - {\sqrt{97}}
\right) }\over 4}$&${{-5\,\left( 17 + {\sqrt{97}} \right) }\over
{16}}$&${{-5\,\left( 31 - {\sqrt{97}} \right) }\over 8}$&$0$&$0$\\
 $  7$&$1$&$-6$&${{41 - {\sqrt{97}}}\over 4}$&${{3\,\left( 31 + {\sqrt{97}}
\right) }\over 4}$&${{-5\,\left( 17 - {\sqrt{97}} \right) }\over
{16}}$&${{-5\,\left( 31 + {\sqrt{97}} \right) }\over 8}$&$0$&$0$\\
 $  8$&$0$&$1$&$-3$&$0$&${{601 + {\sqrt{4801}}}\over {264}}$&${{59 -
{\sqrt{4801}}}\over {44}}$&$0$&$0$\\
 $  9$&$0$&$1$&$-3$&$0$&${{601 - {\sqrt{4801}}}\over {264}}$&${{59 +
{\sqrt{4801}}}\over {44}}$&$0$&$0$\\
 $ 10$&$0$&$1$&$-5$&$0$&${{379 + {\sqrt{1081}}}\over {48}}$&${{61 -
{\sqrt{1081}}}\over 8}$&${{-\left( 159 + {\sqrt{1081}} \right) }\over
{40}}$&${{-\left( 61 - {\sqrt{1081}} \right) }\over 8}$\\
 $ 11$&$0$&$1$&$-5$&$0$&${{379 - {\sqrt{1081}}}\over {48}}$&${{61 +
{\sqrt{1081}}}\over 8}$&${{-\left( 159 - {\sqrt{1081}} \right) }\over
{40}}$&${{-\left( 61 + {\sqrt{1081}} \right) }\over 8}$\\
\end{tabular}
\end{table}

This deviation at intermediate $Q^2$ is found to be of  order $0.9$
for ${\cal O}(l) \le 3$ and increases exponentially for higher orders.
Since in the ansatz of Ref.~\onlinecite{EHG92} $\gamma_0 \ge \gamma_n \ge
\gamma_9$, one can approximate
\begin{equation}
{\cal M}_l(Q^2) \mapsto \tilde{\cal M}_l(Q^2) =
{\left(
{\hat{B}_l(\mu^2)\,\left(\frac{\alpha_S(Q^2)}{\alpha_S(\mu^2)}
\right)^{\gamma_l}} \right) }
{ /}
{\left(
{\left(\frac{\alpha_S(Q^2)}{\alpha_S(\mu^2)}\right)^{\bar{\gamma}}
\sum_{n,k} B_n(\mu^2) c_k F_{nk}^l} \right)}.
\end{equation}
\vspace{-0.9 true cm}
\begin{center}
\mbox{
\begin{minipage}{7.0 true cm}
\begin{figure}
\bild{symetric.ps}{6 true cm}{10}{0}{595}{420}
\caption{The eigenvalues of the evolution equation (for $S_n=1$) vs.
the corresponding order ${\cal O}(n)=M$ (solid line).} \label{fig1}
\end{figure}
\end{minipage} \ \
\begin{minipage}{7.9 true cm}
\vspace{-18 pt}
\begin{figure}
\bild{mtilde.ps}{6.5 true cm}{-15}{370}{560}{825}
\caption{Deviation measure $\tilde{\cal M}_l(Q^2 = 30$\ $ GeV^2/c^2)$
for the two
extreme cases $\bar{\gamma}=\gamma_0$ and $\bar{\gamma}=\gamma_9$.}
\label{fig2}
\end{figure}
\end{minipage}
}
\end{center}
\par

Using the power-law behavior for $\gamma_l$, Fig.~\ref{fig2}
shows the exponetially increasing deviation of the approximation of
Ref.~\onlinecite{EHG92} compared to the $Q^2$-scaling behavior
based on the renormalization group equation.
This comparison shows that it is important to
project the nucleon distribution amplitude
$\Phi_N$, as in the case of mesons~\cite{HS91}, on the eigenfunctions of
the evolution equation. For this purpose, powerful tools have been
developed and the basis of eigenfunctions $\tilde{\Phi}_n$ has been
extended up to polynomial order 9. Using analytical and numerical
algorithms developed in Refs.~\onlinecite{Ber93,BS93} the calculation
of higher order eigenfunctions up to any desired precision is shown
to be possible.  \par

{\it This work was supported in part by the Deutsche
Forschungsgemeinschaft and the COSY-J\"ulich project. \par} \par
\vspace{0.2 true cm}

\end{document}